\documentclass[prd,superscriptaddress,amsfonts,amssymb,amsmath,showpacs,twocolumn]{revtex4-1}
\usepackage{textcomp}
\usepackage{eurosym}
\usepackage{amsfonts}
\usepackage{array}
\usepackage{amsthm}
\usepackage{bm}
\usepackage[colorinlistoftodos]{todonotes}
\usepackage{mathpazo}
\usepackage{supertabular}
\usepackage{subfig}
\usepackage{amssymb}
\usepackage{eurosym}
\usepackage{amsmath}
\usepackage{epsfig}
\usepackage{graphics}
\usepackage{color}
\usepackage{graphicx}
\usepackage[font={footnotesize,it}]{caption}
\usepackage[colorlinks=true,
            linkcolor=blue,
            urlcolor=blue,
            citecolor=green]{hyperref}
\begin{document}
\title{Conical Morris-Thorne Wormholes with a Global Monopole Charge}
\author{Kimet Jusufi}
\email{kimet.jusufi@unite.edu.mk}
\affiliation{Physics Department, State University of Tetovo, Ilinden Street nn, 1200,
Tetovo, Macedonia}
\affiliation{Institute of Physics, Faculty of Natural Sciences and Mathematics, Ss. Cyril and Methodius University, Arhimedova 3, 1000 Skopje, Macedonia}

\begin{abstract}
In this paper we have established an asymptotically conical Morris-Thorne wormhole solution supported by anisotropic matter fluid and a global monopole charge
in the framework of a $1+3$ dimensional gravity minimally coupled to a triplet of scalar fields $\phi^a$, resulting from the breaking of a global $O(3)$ symmetry. For the anisotropic matter fluid we have considered the equation of state (EoS) given by $\mathcal{P}_r=\omega \rho$, with a consequence $\omega<-1$, implying a so-called phantom energy at the throat of the wormhole which violates the energy conditions. In addition, we study the weak gravitational lensing effect using the Gauss-Bonnet theorem (GBT) applied to the wormhole optical geometry. We show that the total deflection angle consists of a term given by  $4\pi^2 \eta^2 $, which is independent from the impact parameter $b$, and an additional term which depends on the radius of the wormhole throat $b_0$ as well as the dimensionless constant $\zeta$.
\end{abstract}
\maketitle

\section{Introduction}
Wormholes are associated with the amazing spacetime topology of connecting two spacetime geometries located in different regions of the universe or different universes. They are solutions of the Einstein field equations, historically, the first step toward the concept of wormholes was made by Flamm \cite{Flamm}, later on a new spin was put forward by Einstein and Rosen \cite{Einstein}. It is interesting to note that Einstein and Rosen proposed a geometric model for elementary particles, such as the electron, in terms of the Einstein-Rosen bridge (ERB). However, this model it turns out to be unsuccessful, moreover the ERB was shown to be unstable \cite{FullerWheeler,whel,Wheeler,Wheeler1,Ellis,Ellis1}.

Traversable wormholes were studied extensively in the past by several authors, notably  Ellis \cite{Ellis,Ellis1} and Bronnikov \cite{br1} studied exact traversable wormhole solutions with a phantom scalar, while few years later different wormhole models were discussed by Clement \cite{clm}, followed by the seminal paper by Morris and Thorne \cite{Morris}. Afterwards, Visser developed the concept of thin-shell wormholes \cite{Visser1}. Based on physical grounds, it is well known that all the matter in our universe obeys certain energy conditions, in this context, as we shall see the existence of wormholes is problematic. In particular, the geometry of TW requires a spacial kind of exotic matter concentrated at the wormhole throat (to keep the spacetime region open at the throat). In other words, this kind of matter violates the energy conditions, such as the null energy condition (NEC) \cite{Visser1}. It is speculated that such a matter can exists in the context of quantum field theory. The second problem is related to the stability of the wormholes. Given the wormhole spacetime geometry, one way to check the stability analyses is the linear perturbation method around the wormhole throat proposed by Visser and Poisson \cite{visser2}.  Wormholes have been studied in the framework of different gravity theories, for example the rotating traversable wormhole solution found by Teo \cite{Teo}, spinning wormholes in scalar-tensor theory \cite{kunz1}, wormholes with phantom energy \cite{lobo0}, wormholes in Gravity's Rainbow \cite{lobo1}, traversable Lorentzian wormhole with a cosmological constant \cite{lobo2}, wormholes in Einstein-Cartan theory \cite{branikov1}, wormholes in Eddington-inspired Born-Infeld gravity \cite{r1,r2,r3}, wormholes with different scalar fields and charged wormholes \cite{kunz2,barcelo,kim,branikov0,habib,barcelo,kunz2,branikov,jamil}, wormholes from cosmic strings \cite{clement1}, wormholes by GUTs in the early universe \cite{nojiri}, wormholes in $f(R,T)$ gravity \cite{moraes} and recently \cite{hardi,myrzakulov,sar}. Recently, extensive studies have been conducted by different authors  related to the thin-shell wormhole approach  \cite{rahaman,lobo3,lobo4,farook1,eiroa,ali,kimet,Jusufi:2016eav,Ovgun:2016ujt,Halilsoy:2013iza}.

Topological defects are interesting objects predicted to exist by particle physics due to the phase transition mechanism in the early universe \cite{Kibble}. One particular example of topological defects is the global monopole, a spherically symmetric object resulting from the self-coupling triplet of scalar fields $\phi^a$ which undergoes a spontaneous breaking of global $O(3)$ gauge symmetry down to $U(1)$. The spacetime metric describing the global monopole has been studied in many papers including \cite{vilenkin,vilenkin1,narash,Bertrand}. In this latter we provide a new Morris-Thorne wormhole solution with anisotropic fluid and a global monopole charge in $1+3$ gravity theory minimally coupled to a triplet of scalar fields. The deflection of light by black holes and wormholes has attracted great interest, in this context the necessary methodology can be found in the papers by Bozza \cite{bozza1,bozza2,bozza3,bozza4}, Perlick \textit{et al.} \cite{perlick1,perlick2,perlick3,perlick4}, and Tsukamoto \textit{et al.} \cite{t1,t2,t3,t4,t5,t6}. For some recent works concerning the strong/weak lensing see also \cite{wh0,asada,potopov,abe,strong1,nandi,ab,mishra,f2,kuh,Sharif:2015qfa,Hashemi:2015axa,Sajadi:2016hko,Pradhan:2016qxa,Lukmanova:2016czn,Nandi:2016uzg}. While for an alternative method to study gravitational lensing via GBT see the Refs. \cite{GibbonsWerner1,K1,K2,K3,K4,K5}.

This paper has the following organization. In Sec. 2, we deduce  the metric for a static and spherically symmetric Morris-Thorne wormhole with a global monopole charge. In Sec. 3, we study the weak gravitational lensing applying the Gauss-Bonnet theorem. In Sec. 4, we draw our conclusions.

\section{Morris-Thorne Wormhole with a Global Monopole charge}
We start by writing the $3 + 1$ −dimensional action without a cosmological constant
minimally coupled to a scalar field with  matter fields, in units $c=G=1$ given by
\begin{equation}
S=\int \left(\frac{\mathfrak{R}}{2 \kappa}+\mathcal{L}\right) \sqrt{-g}\,\mathrm{d}^4x+S_m\label{1}
\end{equation}
in which $\kappa= 8\pi$. The Lagrangian density describing a self-coupling scalar triplet $\phi^{a}$ is given by \cite{vilenkin}
\begin{equation}
\mathcal{L}=-\frac{1}{2}\sum_a g^{\mu\nu}\partial_{\mu}\phi^{a} \partial_{\nu}\phi^{a}-\frac{\lambda}{4}\left(\phi^{2}-\eta^{2}\right)^{2},\label{2}
\end{equation}
with $a=1, 2, 3$, while $\lambda$ is the self-interaction term, $\eta$ is the scale of a gauge-symmetry breaking. The field configuration describing a monopole is
\begin{equation}
\phi^{a}=\frac{\eta f(r) x^{a}}{r},
\end{equation}
in which
\begin{equation}
x^{a}=\left\lbrace r \sin\theta \, \cos\varphi, r \sin\theta \,\sin\varphi,r \cos\theta \,\right\rbrace,
\end{equation}
such that $\sum_a x^{a}x^{a}=r^{2}$.  Next, we consider a static and spherically symmetric Morris-Thorne traversable wormhole in the Schwarzschild coordinates given by \cite{Morris} 
\begin{equation}
\mathrm{d}s^{2}=-e^{2\Phi (r)}\mathrm{d}t^{2}+\frac{\mathrm{d}r^{2}}{1-\frac{b(r)}{r}}+r^{2}\left(
\mathrm{d}\theta ^{2}+\sin ^{2}\theta \mathrm{d}\varphi ^{2}\right),  \label{5}
\end{equation}
in which $\Phi (r)$ and $b(r)$ are the redshift and shape
functions, respectively. In the wormhole geometry, the redshift
function $\Phi (r)$ should be finite in order to avoid the
formation of an event horizon. Moreover, the shape function $b(r)$ determines the wormhole geometry, with the following condition $b(r_{0})=r_{0}$, in which $r_{0}$ is the radius of the wormhole throat. Consequently, the shape function must satisfy the flaring-out condition \cite{lobo0}: 
\begin{equation}
\frac{b(r)-rb^{\prime }(r)}{b^{2}(r)}>0, 
\end{equation}%
in which $b^{\prime }(r)=\frac{db}{dr}<1$ must hold at the throat of the
wormhole. The Lagrangian density in terms of $f$ reads
\begin{equation}
\mathcal{L}=-\left(1-\frac{b(r)}{r}\right)\frac{\eta^{2}(f^{\prime})^{2}}{2}-\frac{\eta^{2}f^{2}}{r^{2}}-\frac{\lambda \eta^{4}}{4}\left(f^{2}-1\right)^{2}.
\end{equation}

On the other hand the Euler-Lagrange equation for the field $f$ gives
\begin{eqnarray}\notag
\left(1-\frac{b}{r}\right)f^{\prime\prime}+&f^{\prime}&\left[\left(1-\frac{b(r)}{r}\right)\frac{2}{r}+\frac{1}{2}\left(\frac{b-b'r}{r^2}\right)\right]\\
&-& f\left[\frac{2}{r^{2}}+\lambda \eta^{2} \left(f^{2}-1\right)\right]=0.\label{8}
\end{eqnarray}

The energy momentum tensor from the Lagrangian density \eqref{2} is found to be
\begin{equation}
\bar{T}_{\mu\nu}=\partial_{\mu}\phi^{a}\partial_{\nu}\phi^{a}-\frac{1}{2}g_{\mu\nu}g^{\rho \sigma}\partial_{\rho}\phi^{a}\partial_{\sigma}\phi^{a}-\frac{g_{\mu\nu} \lambda}{4}\left(\phi^{a}\phi^{a}-\eta^{2}\right)^{2}.
\end{equation}

Using the last equation, the energy-momentum components are given as follows
\begin{equation}
\bar{T}_{t}^{t}=-\eta^{2}\left[\frac{f^{2}}{r^{2}}+\left(1-\frac{b}{r}\right)\frac{(f^{\prime})^{2}}{2}+\frac{\lambda \eta^{2}}{4}(f^{2}-1)^{2}\right],
\end{equation}
\begin{equation}
\bar{T}_{r}^{r}=-\eta^{2}\left[\frac{f^{2}}{r^{2}}-\left(1-\frac{b}{r}\right)\frac{(f^{\prime})^{2}}{2}+\frac{\lambda \eta^{2}}{4}(f^{2}-1)^{2}\right],
\end{equation}
\begin{equation}
\bar{T}_{\theta}^{\theta}=\bar{T}^{\varphi}_{\varphi}=-\eta^{2}\left[\left(1-\frac{b}{r}\right)\frac{(f^{\prime})^{2}}{2 }+\frac{\lambda \eta^{2}}{4}(f^{2}-1)^{2}\right].
\end{equation}

It turns out that Eq. \eqref{8} cannot be solved exactly, however it suffices  to set $f(r)\to 1$ outside the wormhole.  Consequently, the energy-momentum components reduces to 
\begin{equation}
\bar{T}_{t}^{t} = \bar{T}_{r}^{r} \simeq - \frac{\eta^{2}}{r^{2}},\,\,\,\,\bar{T}_{\theta}^{\theta} = \bar{T}_{\varphi}^{\varphi} \simeq 0 .
\end{equation}

On the other hand Einstein's field equations (EFE) reads
\begin{equation}
G_{\mu \nu }=R_{\mu \nu }-\frac{1}{2}g_{\mu \nu }R=8\pi \mathcal{T}_{\mu \nu },
\label{68}
\end{equation}%
where $\mathcal{T}_{\mu \nu }$ is the total energy-momentum tensor which can be written as a sum of the matter fluid part and the matter fields
\begin{equation}
\mathcal{T}_{\mu \nu }=T_{\mu \nu }^{(0)}+\bar{T}_{\mu \nu }.  \label{15}
\end{equation}%

For the matter fluid we shall consider an anisotropic fluid with the following energy-momentum tensor components 
\begin{equation}
{{T^{\mu }}_{\nu }}^{(0)}=\left( -\rho ,\mathcal{P}_{r},\mathcal{P}_{\theta },%
\mathcal{P}_{\varphi }\right).  \label{16}
\end{equation}%

Einstein tensor components for the generic wormhole metric \eqref{5} gives
\begin{eqnarray}
G_{t}^{t} &=&-\frac{b^{\prime }(r)}{r^{2}},  \notag \\
G_{r}^{r} &=&-\frac{b(r)}{r^{3}}+2\left( 1-\frac{b(r)}{r}\right) \frac{\Phi
^{\prime }}{r},  \notag \\
G_{\theta }^{{\theta }} &=&\left( 1-\frac{b(r)}{r}\right) \Big[\Phi ^{\prime
\prime }+(\Phi ^{\prime })^{2}-\frac{b^{\prime }r-b}{2r(r-b)}\Phi ^{\prime }
\notag \\
&-&\frac{b^{\prime }r-b}{2r^{2}(r-b)}+\frac{\Phi ^{\prime }}{r%
}\Big],  \notag \\
G_{\varphi }^{{\varphi }} &=&G_{\theta }^{{\theta }}.  \label{73n}
\end{eqnarray}

The energy-momentum components yields
\begin{eqnarray}
\rho (r) &=&\frac{1}{8\pi r^{2}}\left[ b^{\prime }(r)-8 \pi \eta^2 \right] ,  \notag \\
\mathcal{P}_{r}(r) &=&\frac{1}{8\pi }\left[ 2\left( 1-\frac{b(r)}{r}\right) 
\frac{\Phi ^{\prime }}{r}-\frac{b(r)}{r^{3}}+\frac{8 \pi \eta^2}{r^2}\right]
,  \notag \\
\mathcal{P}(r) &=&\frac{1}{8\pi }\left( 1-\frac{b(r)}{r}\right) \Big[\Phi
^{\prime \prime }+(\Phi ^{\prime })^{2}-\frac{b^{\prime }r-b}{2r(r-b)}\Phi
^{\prime }  \notag \\
&-&\frac{b^{\prime }r-b}{2r^{2}(r-b)}+\frac{\Phi ^{\prime }}{r%
}\Big].  \label{18}
\end{eqnarray}%
where $\mathcal{P}=\mathcal{P}_{\theta }=\mathcal{P}_{\varphi }$. To simplify the problem, we use the EoS  of the form \cite{lobo0,lobo1,lobo2}
\begin{equation}
\mathcal{P}_{r}=\omega \rho .
\end{equation}

In terms of the equation of state, from Eq. \eqref{18} it is possible to find the following result
\begin{equation}
\frac{b(r)-8\pi \eta^2 r+8 \pi \omega \rho r^3-2r(r-b(r))\Phi'(r)}{r^3}=0.\label{k20}
\end{equation}

Substituting the energy density relation 
\begin{equation}
\rho (r) =\frac{1}{8\pi r^{2}}\left[ b^{\prime }(r)-8 \pi \eta^2 \right],\label{kim21}
\end{equation}
into Eq. \eqref{k20} we find
\begin{equation}
\frac{b'(r)\omega r+b(r)-8\pi \eta^2(\omega+1) r-2r(r-b(r))\Phi'(r)}{r^3}=0.\label{kim22}
\end{equation}

In our setup we shall consider a constant redshift function, namely a wormhole solution with zero tidal force, i.e., $\Phi'=0$, therefore last equation simplifies to
\begin{equation}
b'(r)\omega r+b(r)-8\pi \eta^2(\omega+1) r=0.
\end{equation}

Finally we use the condition $b(r_0)=b_0=r_0$, thus by solving the last differential equation we find the shape function to be
\begin{equation}\label{24}
b(r)=\left(\frac{r_0}{r}\right)^{1/\omega}r_0(1-8\pi \eta^2) +8 \pi \eta^2r.
\end{equation}

One can observe that the wormhole solution is not asymptotically flat by checking the following equation
\begin{equation}
\lim_{r\to \infty} \frac{b(r)}{r}\to \lim_{r\to \infty}\left[\left(\frac{r_0}{r}\right)^{1+\frac{1}{\omega}}(1-8\pi \eta^2)\right] + 8 \pi \eta^2.\label{kim25}
\end{equation}

The first term blows up when $r\to \infty$, since $\omega<-1$. With the help of the shape function the wormhole metric reduces to
\begin{equation}\label{26}
\mathrm{d}s^{2}=-\mathrm{d}t^{2}+\frac{\mathrm{d}r^{2}}{\left(1-8 \pi \eta^2\right)\left[1- \left(\frac{r_0}{r}\right)^{1+\frac{1}{\omega}}\right]}+r^{2}d\Omega^2.  
\end{equation}%

\begin{figure}[h!]
\center
\includegraphics[width=0.45\textwidth]{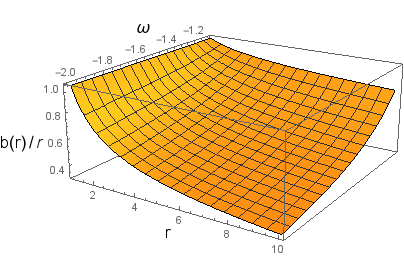}
\caption{{\protect\small \textit{ The figure shows the behavior the shape function $b(r)/r$
as a function of $r$ and $\omega$, for chosen $b_0=1$ and $\eta=10^{-5}$. }}}
\label{f2}
\end{figure}

Note that the constant factor $\exp(2 \Phi)=const$, is absorbed into the re-scaled time coordinate $t$.  To our best knowledge, this metric is reported here for the first time. On the other hand, the metric coefficient $g_{rr}$ diverges at the throat $b(r_0)=r_0$, however this just signals the coordinate singularity. To see this, one can calculate the scalar curvature or the Ricci scalar which is found to be 
\begin{equation}
\mathfrak{R}=\frac{16 \pi \eta^2}{r^2}+(1-8 \pi \eta^2) \frac{2 r_0}{\omega r^2}\left(\frac{r_0}{r}\right)^{1+\frac{1}{\omega}},
\end{equation}
from the last equation we see that the metric is regular at  $r=r_0$. Due to the above coordinate singularity it is convenient to compute the  the proper radial distance which should be a finite quantity
\begin{equation}
l=\pm \int_{r_0}^{r} \frac{\mathrm{d}r'}{\sqrt{1-\frac{b_{\pm}(r')}{r'}}}.
\end{equation}

Using Eq. \eqref{24} we find
\begin{equation}
l(r)=\pm \frac{\Big[r F_1\left(\frac{1}{2},-\frac{\omega+1}{\omega},\frac{1}{\omega +1}, (\frac{r_0}{r})^{\frac{\omega +1}{\omega}}\right)-\frac{r_0 \sqrt{\pi}\Gamma(\frac{1}{\omega +1})}{\Gamma(\frac{1}{2}-\frac{\omega}{\omega +1})}    \Big]}{\sqrt{1-8\pi \eta^2}}
\end{equation}
in which $\pm$ stands for the upper and lower part, respectively. Next, we verify whether the null energy condition (NEC), and weak energy condition (WEC) are satisfied at the throat of the wormhole. As we know WEC is defined by $T_{\mu \nu }U^{\mu }U^{\nu }\geq 0$ i.e., $\rho \geq 0$ and $%
\rho (r)+\mathcal{P}_{r}(r)\geq 0$, where $T_{\mu \nu }$ is the energy
momentum tensor with $U^{\mu }$ being a timelike vector. On the other hand, NEC can be defined by $T_{\mu \nu }k^{\mu }k^{\nu }\geq 0$ i.e., $\rho (r)+%
\mathcal{P}_{r}(r)\geq 0$, with $k^{\mu }$ being a null vector. In this
regard, we have the following energy condition at the throat region: 
\begin{equation}
\rho (r_{0})=\frac{b^{\prime }(r_{0})-8 \pi \eta^2}{8\pi r_{0}^{2}}.  \label{83}
\end{equation}%
Now, using the field equations, one finds the
following relations 
\begin{equation}
\rho (r)+\mathcal{P}_{r}(r)=\frac{1}{8\pi }\Big[\frac{b'(r_0)r_0-b(r_0)}{r_0^3}\Big],
\end{equation}%
considering now the shape function at the throat region we find
 \begin{eqnarray}
\left(\rho+\mathcal{P}_r\right)|_{r=r_0} =- \frac{(1-8\pi \eta^2)(\omega+1)}{ 8\omega \pi r_0^2},
 \end{eqnarray}
this result verifies that matter configuration violates the energy conditions at the throat 
$
\left(\rho+\mathcal{P}_r\right)|_{r=r_0}< 0. 
$ 

Another way to see this is simply by using the flaring-out condition 
\begin{equation}
b'(r_0) = 8 \pi \eta^2-\frac{(1-8 \pi \eta^2)}{\omega }< 1
\end{equation}
which implies $\omega<-1$.  This form of exotic matter with $\omega <-1$, is usually known as a phantom energy. Another important quantity is the ``volume integral quantifier," which basically measures the amount of exotic matter needed for the wormhole defined as follows
\begin{equation}
\mathcal{I}_V =\int\left(\rho(r)+\mathcal{P}_r(r)\right)\mathrm{d}V,
\end{equation}
with the volume element given by $\mathrm{d}V=r^2\sin\theta \mathrm{d}r \mathrm{d}\theta \mathrm{d}\phi$. 

For simplicity, we shall evaluate the volume-integral associated to the phantom energy of our wormhole spacetime \eqref{26} by assuming an arbitrary small region, say $r_0$ to a radius situated at $`a'$, in which the exotic matter is confined. More specifically, by considering our shape function $b(r)$ given by Eq. \eqref{24}, for the amount of exotic matter we find
\begin{eqnarray}
\mathcal{I}_V=\frac{(\omega+1)(1-8 \pi \eta^2)}{2 \omega}(a-r_0).
\end{eqnarray}

For an interesting observation when $a \rightarrow r_0$ then it follows
\begin{equation}
\int{(\rho+\mathcal{P}_r)} \rightarrow 0,
\end{equation}
and thus one may interpret that wormhole can be contracted for with arbitrarily small quantities of ANEC violating matter.

\begin{figure}[h!]
\center
\includegraphics[width=0.45\textwidth]{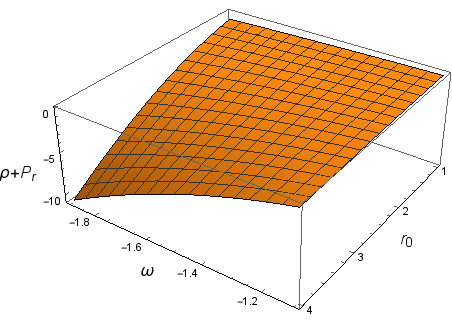}
\caption{{\protect\small \textit{ In this figure we depict the behavior of
$\rho+\mathcal{P}_r$ as a function of $r$ and $\omega$. We have chosen $b_0=1$ and $\eta=10^{-5}$. The energy conditions are violated. }}}
\end{figure}

As we already saw from \eqref{kim25} the first term blows up when $r\to \infty$, since $\omega<-1$. In order to overcome this problem, it is convinient to rewrite the shape function in terms of new dimensionless constants. In particular following Lobo \textit{at al.} \cite{loboasym}, we can consider the following shape function given by
\begin{equation}
\frac{b(r)}{r_0}=a \left[ \left(\frac{r}{r_0}\right)^{\zeta}(1-8 \pi \eta^2)+8 \pi \eta^2 \left(\frac{r}{r_0}\right)\right] +C
\end{equation}
where $a$, $\zeta$, and $C$, are dimensionless constants. Without loss of generality we choose $a=1$, then using $b(r_0)/r_0=1$, we find $C=0$. Furthermore, considering a positive energy density implies $\zeta>0$, while the flaring-out condition imposes an additional constraint at the throat, namely $\zeta <1$. Moreover using the equation of state
at the throat $\mathcal{P}_r(r_0)=\omega \, \rho(r_0)$, we find $\zeta \omega=-1$. On the other hand from Eqs. \eqref{kim21} and \eqref{kim22} we can deduce the following equation
\begin{equation}
\Phi'=\frac{b(r)-8 \pi \eta^2 r+\omega r (b'(r)-8 \pi \eta^2)}{2 r^2 (1-b(r)/r)}.
\end{equation}

To this end using the condition $\zeta \omega=-1$ at $r=r_0$ we find that $\Phi=const$. With this information in hand we can write our wormhole metric as follows
\begin{equation}
\mathrm{d}s^{2}=-\mathrm{d}t^{2}+\frac{\mathrm{d}r^{2}}{\left(1-8 \pi \eta^2\right)\left[1- \left(\frac{b_0}{r}\right)^{1-\zeta}\right]}+r^{2}\mathrm{d} \Omega^2,\label{kim42}
\end{equation}
provided that $\zeta $ is in the range $0<\zeta <1$. Now one can check that 
\begin{equation}
\lim_{r\to \infty} \frac{b(r)}{r}=\lim_{r\to \infty} \left(\frac{r_0}{r}\right)^{1-\zeta}(1-8 \pi \eta^2)+8 \pi \eta^2=8 \pi \eta^2.
\end{equation}
provided $0<\zeta <1$. This equation shows that our wormhole metric \eqref{kim42} is asymptotically conical with a conical deficit angle which is independent of the radial coordinate $r$. Furthermore we can construct the embedding diagrams to visualize the conical wormhole by considering an equatorial slice, $\theta  = \pi/2$ and a fixed moment of time, $t = const$, it follows 
\begin{equation}
\mathrm{d}s^2=\frac{\mathrm{d}r^2}{1-\frac{b(r)}{r}}+r^2 \mathrm{d} \varphi^2.
\end{equation}

On the other hand, we can embed the metric into three-dimensional Euclidean space written in terms of cylindrical coordinates as follows
\begin{equation}
\mathrm{d}s^2=dz^2+\mathrm{d}r^2 + r^2 \mathrm{d}\varphi^2=\left[1+\left(\frac{\mathrm{d}z}{\mathrm{d}r}\right)^2\right]\mathrm{d}r^2+r^2 \mathrm{d} \varphi^2.
\end{equation}

From these equations we can deduce the equation for the embedding surface as follows
\begin{equation}
\frac{\mathrm{d}z}{\mathrm{d}r}=\pm \frac{1}{\sqrt{\left(1-8 \pi \eta^2 \right)\left[ 1-\left(\frac{b_0}{r}\right)^{1-\zeta} \right]}}.
\end{equation}

\begin{figure}
  \center
\includegraphics[width=0.3\textwidth]{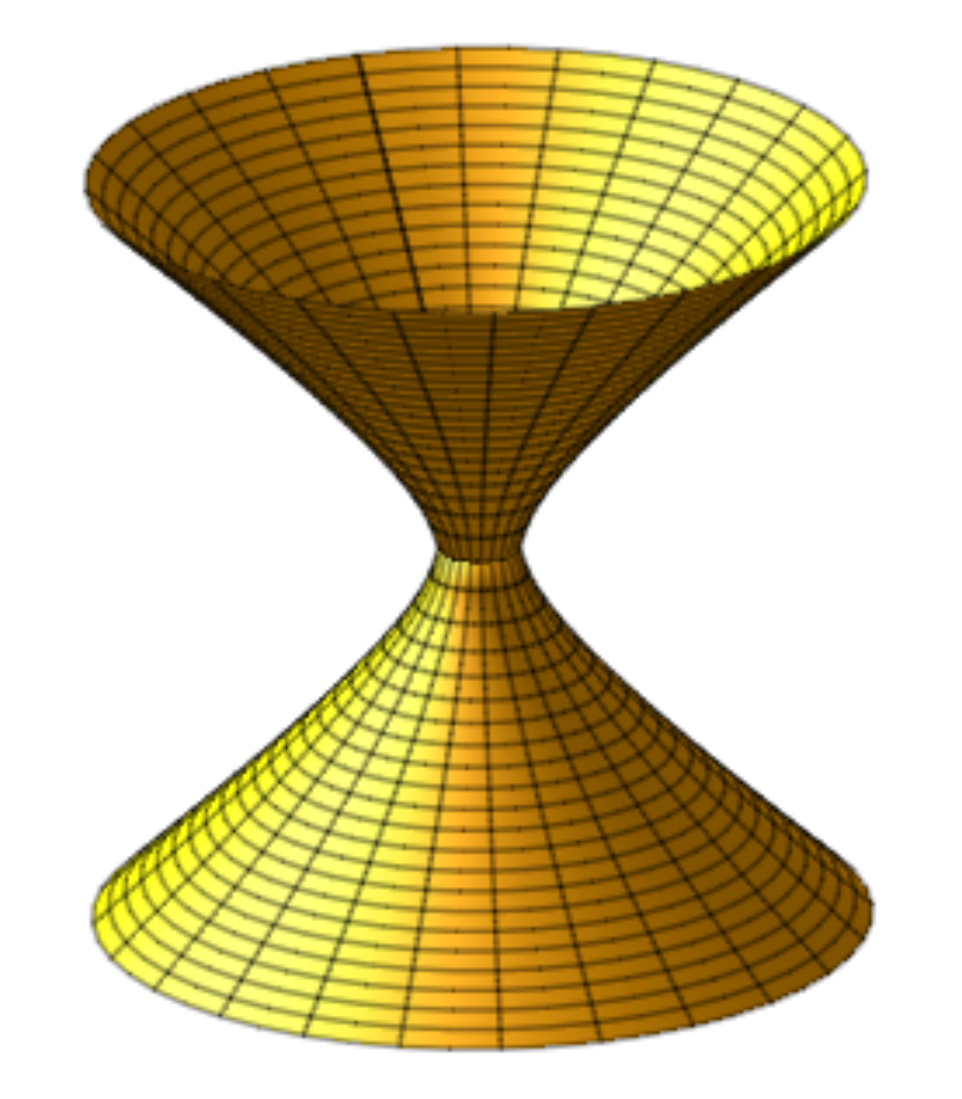}
\caption{{\protect\small \textit{ The embedding diagram of a two-dimensional section
along the equatorial plane with $t = const,$ and $\theta =\pi/2$. To visualize this we plot
$z$ vs. $r$
sweep through a $2 \pi $ rotation around the $z$-axis. We chose $b_0=1$, $\eta=0.01$ and $\zeta=0.5$.}}}\label{conical}
  \end{figure}

Finally we can evaluate this integral numerically for specific parameter values in order to illustrate the conical wormhole shape given in Fig. \ref{conical}.

\section{Gravitational lensing}

We can now proceed to elaborate the gravitational lensing effect in the spacetime of the wormhole metric \eqref{kim42}. The wormhole optical metric can be simply find letting $\mathrm{d}s^2=0$, resulting with
\begin{equation}
\mathrm{d}t^{2}=\frac{\mathrm{d}r^{2}}{\left(1-8 \pi \eta^2\right)\left[1- \left(\frac{b_0}{r}\right)^{1-\zeta}\right]}+r^{2}\mathrm{d} \varphi^2.
\end{equation}

Consequently the optical metric can be written in terms of new coordinates
\begin{equation}
\mathrm{d}t^{2}=h_{ab}\,\mathrm{d}y^{a}\mathrm{d}y^{b}=\mathrm{d}u^2+\mathcal{H}^2(u)\mathrm{d}\varphi
^{2},
\end{equation}
in which we have introduced $\mathcal{H}=r$ and
\begin{equation}
\mathrm{d}u=\frac{\mathrm{d}r}{{\sqrt{\left(1-8 \pi \eta^2 \right)\left(1- \left(\frac{b_0}{r}\right)^{1-\zeta}\right)}}}.
\end{equation}

It is very important to compute first the Gaussian optical curvature (GOC) $\mathcal{K}$ which is defined in terms of the following equation \cite{GibbonsWerner1}
\begin{equation}
\mathcal{K}=-\frac{1}{\mathcal{H}(u)}\left[ \frac{\mathrm{d}r}{\mathrm{d}u}%
\frac{\mathrm{d}}{\mathrm{d}r}\left( \frac{\mathrm{d}r}{\mathrm{d}u}%
\right) \frac{\mathrm{d}\mathcal{H}}{\mathrm{d}r}+\left( \frac{\mathrm{d}r}{\mathrm{d}%
u}\right) ^{2}\frac{\mathrm{d}^{2}\mathcal{H}}{\mathrm{d}r^{2}}\right] .
\end{equation}

Applying this to our optical metric we find
\begin{equation}
\mathcal{K}=-\frac{(1-8 \pi \eta^2)}{2 r^2}\left(\frac{b_0}{r} \right)^{1-\zeta}\left(1-\zeta \right).
\end{equation}

Obviously the GOC is affected by the global monopole charge and the state parameter. Note the important negative sign which is implying the divergence of light rays in the wormhole geometry. But, as we are going to see this is crucial in evaluating the deflection angle which is really a result of a global spacetime topology in terms of the Gauss-Bonnet theorem (GBT). Thus, in our setup we first choose a non-singular domain, or a region outside the light ray noted as $\mathcal{A}_{R}$, with boundaries $\partial 
\mathcal{A}_{R}=\gamma _{h}\cup C_{R}$. Then, the global GBT in terms of the above construction is formulated as follows 
\begin{equation}
\iint\limits_{\mathcal{A}_{R}}\mathcal{K}\,\mathrm{d}\sigma+\oint\limits_{\partial \mathcal{%
A}_{R}}\kappa \,\mathrm{d}t+\sum_{k}\psi _{k}=2\pi \chi (\mathcal{A}_{R}).
\label{10}
\end{equation}

In this equation $\kappa $ is usually known as the geodesic curvature (GC) and basically measures the deviation from the geodesics; $\mathcal{K}$  is the GOC; $\mathrm{d}\sigma$ is the optical surface element; finally $\psi _{k}$ notes the exterior angle at the 
$k^{th}$ vertex. The domain is chosen to be outside of the light
ray implying the  Euler characteristic number to be $\chi (\mathcal{A}_{R})=1$. The
GC is defined via
\begin{equation}
\kappa =h\,\left( \nabla _{\dot{\gamma}}\dot{\gamma},\ddot{\gamma}%
\right),  
\end{equation}%
where we impose the unit speed condition $h(\dot{\gamma},\dot{\gamma})=1$.
For a very large radial coordinate $R\rightarrow \infty $, our two jump angles (at the source $\mathcal{S}$, and observer $\mathcal{O})
$, yields $\psi _{\mathit{O}}+\psi _{\mathit{S}}\rightarrow \pi $ 
\cite{GibbonsWerner1}. Then the GBT simplifies to
\begin{equation}
\iint\limits_{\mathcal{A}_{R}}\mathcal{K}\,\mathrm{d}\sigma+\oint\limits_{C_{R}}\kappa \,%
\mathrm{d}t\overset{{R\rightarrow \infty }}{=}\iint\limits_{\mathcal{A}%
_{\infty }}\mathcal{K}\,\mathrm{d}\sigma+\int\limits_{0}^{\pi +\hat{\alpha}}\mathrm{d}%
\varphi =\pi.  \label{12}
\end{equation}

By definition the GC for $\gamma_{h}$ is zero, hence we are left with a contribution from the curve $C_{R}$ located at a coordinate distance $R$ from the coordinate system chosen at the wormhole center in the equatorial plane. Hence we need to compute 
\begin{equation}
\kappa (C_{R})=|\nabla _{\dot{C}_{R}}\dot{C}_{R}|, 
\end{equation}

In components notation the radial part can be written as
\begin{equation}
\left( \nabla _{\dot{C}_{R}}\dot{C}_{R}\right) ^{r}=\dot{C}_{R}^{\varphi
}\,\left( \partial _{\varphi }\dot{C}_{R}^{r}\right) +\Gamma%
_{\varphi \varphi }^{r(op)}\left( \dot{C}_{R}^{\varphi }\right) ^{2}. 
\end{equation}

With the help of the unit speed condition and after we compute the Christoffel symbol related to our optical metric  in the large coordinate radius $R$ we are left with
\begin{eqnarray}
\lim_{R\rightarrow \infty }\kappa (C_{R}) &=&\lim_{R\rightarrow \infty
}\left\vert \nabla _{\dot{C}_{R}}\dot{C}_{R}\right\vert ,  \notag \\
&\rightarrow &\frac{1}{R} \sqrt{1-8 \pi \eta^2}. 
\end{eqnarray}

Hence, GC is in  fact affected by the monopole charge. To see what this means we write the optical metric in this limit for a constant $R$. We find
\begin{eqnarray}
\lim_{R\rightarrow \infty }\mathrm{d}t \rightarrow R\,\mathrm{d}\varphi.
\end{eqnarray}

Putting the last two equation together we see that $\kappa (C_{R})%
\mathrm{d}t=\sqrt{1-8\pi \eta^2}\mathrm{d}\varphi $. This reflects the conical nature of our wormhole geometry, to put more simply, our optical metric is not asymptotically Euclidean. Using this result from GBT we can express the deflection angle as follows
\begin{equation}
\hat{\alpha}=\pi \left[\frac{1}{\sqrt{1-8 \pi \eta^2}}-1 \right]-\frac{1}{\sqrt{1-8 \pi \eta^2}}\int\limits_{0}^{\pi }\int\limits_{\frac{b}{\sin \varphi }%
}^{\infty }\mathcal{K}\mathrm{d}\sigma. 
\end{equation}

If we used the equation for the light ray $r(\varphi )=b/\sin \varphi $, in which $b$ is the impact parameter, which can be approximated with the closest approach distance from the wormhole in the first order approximation. The surface are is also approximated as
\begin{equation}
\mathrm{d}\sigma= \sqrt{h}\,\mathrm{d}u\,\mathrm{d}\varphi \simeq \frac{r}{\sqrt{1-8\pi \eta^2}}.
\end{equation}

Finally the total deflection angle is found to be
\begin{equation}
\hat{\alpha}=4\pi^2 \eta^2+\left(\frac{b_0}{b}\right)^{1-\zeta}\frac{\sqrt{\pi}\,\Gamma\left(1-\frac{\zeta}{2}\right)}{2\, \Gamma\left(\frac{3-\zeta}{2}\right) }.
\end{equation}

We can recast our wormhole metric \eqref{26} in a different form. In particular if we introduce the coordinate transformations
\begin{equation}\label{53}
\mathcal{R}\to \frac{r}{\sqrt{1-8\pi \eta^2}},
\end{equation}
and
\begin{equation}\label{54}
\mathcal{B}_0\to \frac{b_0}{\sqrt{1-8\pi \eta^2}}.
\end{equation}

Taking into the consideration the above transformations the wormhole metric reduces to 
\begin{equation}\label{55}
\mathrm{d}s^{2}=-\mathrm{d}t^{2}+\frac{\mathrm{d}\mathcal{R}^{2}}{1- \left(\frac{\mathcal{B}_0}{\mathcal{R}}\right)^{1-\zeta}}+\left(1-8\pi \eta^2 \right) \mathcal{R}^{2}\mathrm{d}\Omega^2.  
\end{equation}

One can show that the deflection angle remains invariant under the coordinate transformations \eqref{53}-\eqref{54}. In a similar fashion, we can apply the following substitutions $\mathcal{H}=\mathcal{R} \sqrt{1-8 \pi \eta^2}$, and
\begin{equation}
\mathrm{d}u=\frac{\mathrm{d}\mathcal{R}}{{\sqrt{1- \left(\frac{\mathcal{B}_0}{\mathcal{R}}\right)^{1-\zeta}}}}.
\end{equation}

Then, for the GOP in this case it is not difficult to find that
\begin{equation}
\mathcal{K}=-\frac{\left(1-\zeta \right)}{2\mathcal{R}^2}\left(\frac{\mathcal{B}_0}{\mathcal{R}} \right)^{1-\zeta}.
\end{equation}

In the limit $R\rightarrow \infty$, GC yields
\begin{eqnarray}
\lim_{R\rightarrow \infty }\kappa (C_{R}) &=&\lim_{R\rightarrow \infty
}\left\vert \nabla _{\dot{C}_{R}}\dot{C}_{R}\right\vert ,  \notag \\
&\rightarrow &\frac{1}{R},
\end{eqnarray}
but 
\begin{eqnarray}
\lim_{R\rightarrow \infty }\mathrm{d}t \rightarrow R\, \sqrt{1-8 \pi \eta^2}\,\mathrm{d}\varphi.
\end{eqnarray}

Although GC is independent by $\eta$, we see that $\mathrm{d}t$ is affected by $\eta$. However, we end up with the same result  $\kappa (C_{R})
\mathrm{d}t=\sqrt{1-8\pi \eta^2}\mathrm{d}\varphi $. The equation for the light ray this time can be choosen as $\mathcal{R}=\mathcal{B}/\sin \varphi$, resulting with a similar expression 
\begin{equation}
\hat{\alpha}=\pi \left[\frac{1}{\sqrt{1-8 \pi \eta^2}}-1 \right]-\frac{1}{\sqrt{1-8 \pi \eta^2}}\int\limits_{0}^{\pi }\int\limits_{\frac{\mathcal{B}}{\sin \varphi }%
}^{\infty }\mathcal{K}\mathrm{d}\sigma.  \label{17n}
\end{equation}

Solving this integral we can approximate the solution to be
\begin{equation}
\hat{\alpha}=4\pi^2 \eta^2+\left(\frac{\mathcal{B}_0}{\mathcal{B}}\right)^{1-\zeta}\frac{\sqrt{\pi}\,\Gamma\left(1-\frac{\zeta}{2}\right)}{2\, \Gamma\left(\frac{3-\zeta}{2}\right) }.
\end{equation}

From the equations of the light rays we deduce that the impact parameters should be related with 
\begin{equation}
\mathcal{B}\to \frac{b}{\sqrt{1-8\pi \eta^2}},
\end{equation}
yielding the ratio 
\begin{equation}
\frac{\mathcal{B}_0}{\mathcal{B}}\to \frac{b_0}{b}.
\end{equation}

Thus, we showed that the final expression for the deflection angle remains invariant under the coordinate transformations \eqref{53}-\eqref{54}. For an important observation we can compare out result with two special case. Firstly, we note that the metric \eqref{55} reduces to the point like global monopole metric by letting $\mathcal{B}_0=0$, thus
\begin{equation}
\mathrm{d}s^{2}=-\mathrm{d}t^{2}+\mathrm{d}\mathcal{R}^{2}+\left(1-8\pi \eta^2 \right)\mathcal{R}^{2}\mathrm{d}\Omega^2.  
\end{equation}

The deflection angle due to the point like global monopole is given by $4\pi^2 \eta^2$ (see, for example \cite{K5}). It is clear that due to the geometric contribution related to the wormhole thoruat, the light bending is stronger in the wormhole case compared to the point-like global monopole case.\\

\begin{figure}[h!]
\center
\includegraphics[width=0.45\textwidth]{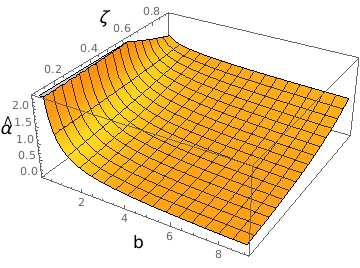}
\caption{{\protect\small \textit{ The figure shows the deflection angle as a function of the impact parameter $b$ and $\zeta$, for chosen $b_0=1$ and $\eta=10^{-5}$. }}}
\label{f3}
\end{figure}

\section{Conclusion}
In this paper, we have found an asymptotically conical Morris-Thorne wormhole supported by anisotropic matter fluid and a triplet of scalar fields $\phi^a$  minimally coupled to a $1+3$ dimensional gravity. For the anisotropic fluid we have used EoS of the form $\mathcal{P}_r=\omega \rho$, resulting with a phantom energy described by the relation $\omega<-1$.
Our phantom wormhole solution is characterized by a solid angle deficit due to the global conical geometry reveling  interesting observational effects such as the gravitational lensing. Introducing a new dimensionless constant $\zeta$ we have shown that our wormhole metric is not asymptotically flat, namely $b(r)/r \to 8 \pi \eta^2 $, when $r \to \infty$. We have also studied the deflection of light, more specifically a detailed analysis using GBT revealed the following result for the deflection angle 
\begin{equation}\nonumber
\hat{\alpha}=4\pi^2 \eta^2+\left(\frac{b_0}{b}\right)^{1-\zeta}\frac{\sqrt{\pi}\,\Gamma\left(1-\frac{\zeta}{2}\right)}{2\, \Gamma\left(\frac{3-\zeta}{2}\right) }.
\end{equation}

Clearly, the first term $4\pi^2 \eta^2 $, is independent of the impact parameter $b$, while the second term is a product of a function written in terms of the throat of the wormhole $b_0/b$, and the Gamma functions depending on the dimensionless constant $\zeta$.  It is worth noting that we have performed our analysis in two different spacetime metrics. In both cases we find the same result hence the deflection angle is form-invariant under coordinate transformations.  Finally we pointed out that the gravitational lensing effect is stronger in the wormhole geometry case compared to the point like global monopole geometry.

\end{document}